\documentclass[%
reprint,
amsmath,amssymb,
aps,
pra,
floatfix
]{revtex4-2}
\usepackage[utf8]{inputenc}
\usepackage[T1]{fontenc}
\usepackage{xcolor}
\usepackage{colortbl}
\usepackage{rotating}
\usepackage{multirow}
\usepackage{amsmath, bm}
\usepackage{natbib}
\usepackage{hyperref}
\setcitestyle{super}

\newcommand{\ket}[1]{\left|#1\right\rangle}
\newcommand{\bra}[1]{\left\langle#1\right|}

\begin{document}

\title{Identification of quantum entanglement with Siamese convolutional NNs and semi-supervised learning}

\author{Jarosław Paw\l{}owski}
\email[]{jaroslaw.pawlowski@pwr.edu.pl}
\affiliation{Institute of Theoretical Physics, 
Wroc{\l}aw University of Science and Technology,
Wybrze\.{z}e Wyspia\'{n}skiego 27,
50-370 Wroc{\l}aw, Poland}
\author{Mateusz Krawczyk}
\affiliation{Institute of Theoretical Physics, 
Wroc{\l}aw University of Science and Technology,
Wybrze\.{z}e Wyspia\'{n}skiego 27,
50-370 Wroc{\l}aw, Poland}


\begin{abstract}
Quantum entanglement is a fundamental property commonly used in various quantum information protocols and algorithms. Nonetheless, the problem of identifying entanglement has still not reached a general solution for systems larger than $2\times3$. 
In this study, we use deep convolutional NNs, a type of supervised machine learning, to identify quantum entanglement for any bipartition in a 3-qubit system. 
We demonstrate that training the model on synthetically generated datasets of random density matrices excluding challenging positive-under-partial-transposition entangled states (PPTES), which cannot be identified (and correctly labeled) in general, leads to good model accuracy even for PPTES states, that were outside the training data.
Our aim is to enhance the model's generalization on PPTES. By applying entanglement-preserving symmetry operations through a triple Siamese network trained in a semi-supervised manner, we improve the model’s accuracy and ability to recognize PPTES. Moreover, by constructing an ensemble of Siamese models, even better generalization is observed, in analogy with the idea of finding separate types of entanglement witnesses for different classes of states.

\end{abstract}

\flushbottom
\maketitle
%
%
\thispagestyle{empty}

\section*{Introduction}

Modern deep learning (DL) architectures, that use multilayer neural networks (NNs), have enabled unprecedented achievements in various domains like computer vision or natural language processing.
Convolutional neural networks (CNNs) with many hidden layers and complex network structures are extremely powerful in feature learning enabling automatic extraction of the most optimal features.
Deep CNNs easily outperform classical algorithms in image classification\cite{alexnet}, object detection\cite{faster, yolo}, or face recognition tasks\cite{facenet}.
In physics, one natural application of DL involves the study of quantum many-body systems\cite{troyer,gao2017,carleo2018,tibaldi}, where the extreme complexity of many-body states often makes theoretical analysis intractable. Nonetheless, employment of machine learning (ML) in physics for entangled state representations is typically focused on more traditional architectures such as Boltzmann machines\cite{carleo2018,rbm} or fully-connected (FC) NNs\cite{ma2018}. However, there are recent signals that deep convolutional and recurrent networks can better represent highly entangled quantum systems~\cite{levine2019,carrasquilla2021}. Therefore, it is worth trying more modern architectures, which are closer to the state-of-the-art approaches in DL, to study entanglement. 

Entanglement is a fundamental feature of quantum physics, where the correlation between subsystems (i.e. particles) cannot be described within a \textit{local} classical model. In quantum information theory, entanglement is regarded as an important resource in achieving various tasks, such as quantum computation, cryptography, and teleportation. Simultaneously, the problem of identifying entanglement has not reached a general solution for systems larger than $2\otimes3$.~\cite{zyczkowski2017, Plenio2014} There exist analytic criteria for separability of a given mixed state which are either sufficient and necessary, but not practically usable (require minimization); or easy to use, but not conclusive~\cite{zyczkowski2017}. One of the criterion of separability, which can be considered is the PPT (positive under partial transpose) criterion\cite{horodecki1997}. It is based on the idea, that having a density matrix partially transposed for a given bipartition, one can determine separability by simply looking at the eigenvalues of this matrix. Namely, if all eigenvalues are positive, then the state is separable in the given bipartition. However, in the case of systems larger than two qubits it is only a necessary condition of separability -- there are no separable states, which have negative eigenvalues, but it is not a sufficient condition -- there exist mixed states, which are entangled although they are PPT. Such states are called PPT entangled states (PPTES). Similarly, all other operationally feasible analytic criteria provide partial solutions only. 

An interesting idea in this area is to implement NNs on a quantum computer~\cite{beer2020,yang2020} and thus create a bridge between DL and quantum computing\cite{schatzki2021}, leading to exponential speedup in training deep NNs\cite{yang2020}. Recently discovered quantum convolutional NNs~\cite{cong2019} allow their efficient training and implementation on realistic quantum devices. Such networks can be used for efficient entanglement detection in multi-qubit systems~\cite{qiu2019}, as well as for solving different classes of quantum many-body problems such as quantum phase recognition~\cite{cong2019}. ML can also be applied to an important problem in quantum algorithm design, the automatic optimization of large unitary matrices into sequences of two-qubit gates~\cite{banchi2016}.
Entanglement can also assist ML image classifiers by compressing images and thus reducing computational resources needed\cite{Lewenstein2021}. 

\begin{figure}[!b]
    \centering
    \includegraphics[scale=.45]{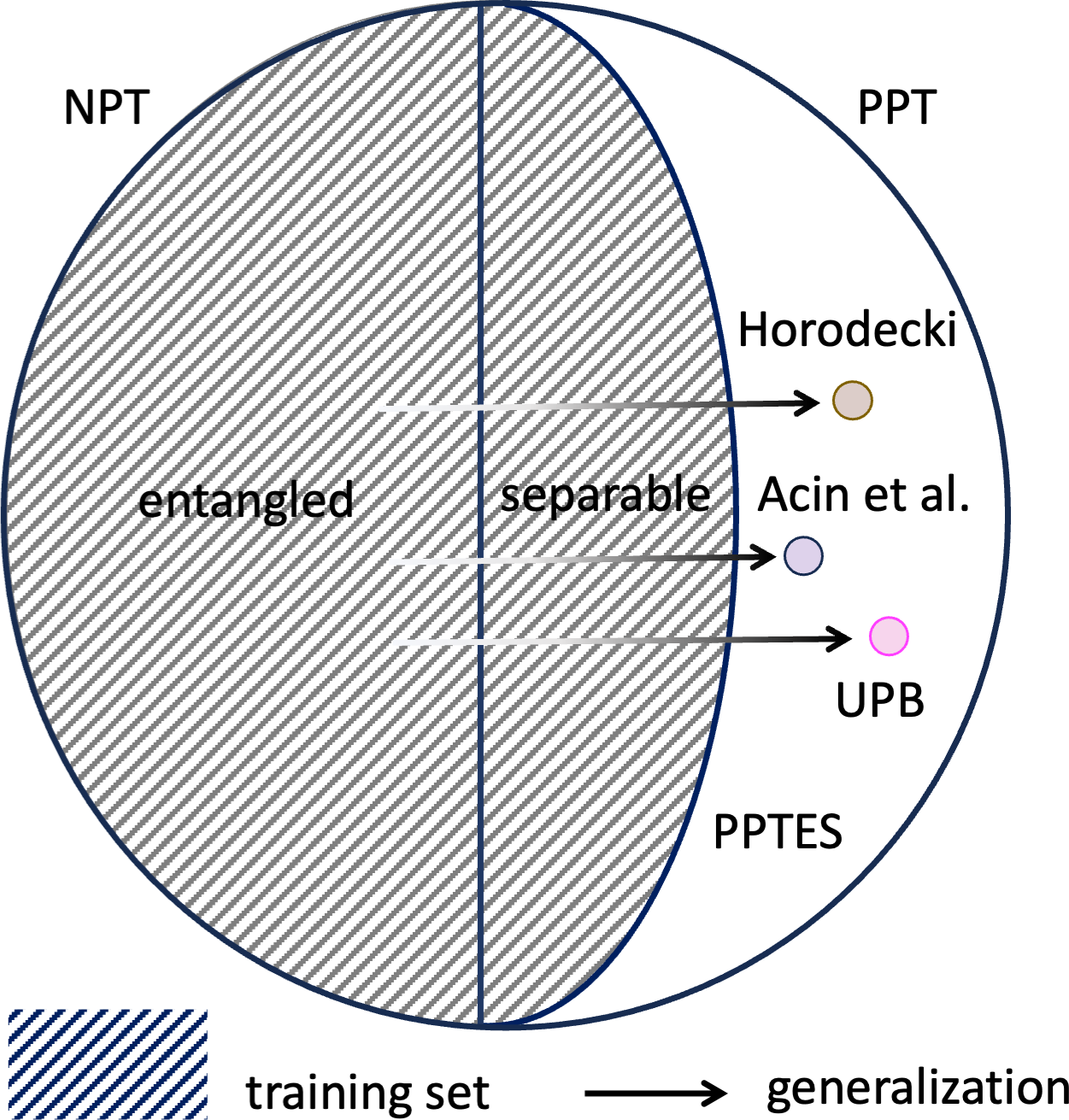}
    \caption{Idea of the supervised training with the training set composed of states we know whether they are entangled or not (\textit{verified} strategy). The goal is to build a model capable of generalizing to PPTES that are hard to recognize by analytical methods.}
    \label{fig:generalization}
\end{figure}

Proposed studies will attempt to identify entanglement using modern DL architectures and confront it with traditional analytic
entanglement metrics. The first applications of advanced DL architectures to such problems are already appearing, both with the utilization of unsupervised\cite{Chen2021} and supervised manner\cite{naema2023, urena2023}.  
In this work, we test a combined approach, which allows us to leverage the strengths and limit the weaknesses of both techniques. Specifically, a lack of appropriate conclusive criterion, which could be used as a label, is a huge problem for supervised learning, and susceptibility to mode collapsing of unsupervised models may lead to incomplete coverage of the desired space, and therefore misclassification of omitted states. Simultaneously, we aim to solve the more general problem of precise entanglement identification by predicting entanglement for each bipartition in a system. In this sense, our model will predict a vector of probabilities that identifies entanglement distribution in the system, rather than just a single entanglement metric.
Moreover, Chen et al.~\cite{Chen2021} validate their method using PPT criterion only. 
Our method tries to deal with more general quantum systems in classifying PPTES, i.e. in situations where the PPT criterion fails, and one cannot obtain reliable labels for supervised learning, testing more advanced NN architectures than simple multi-layer perception (MLP) networks\cite{ma2018}. We also obtain better results than using standard supervised trained NNs\cite{naema2023, urena2023} and classical, non-neural ML algorithms\cite{goes2021,lu2018}.
Therefore our main goal is to build a semi-supervised machine learning system, composed mainly of CNNs, that trained on verified states only can generalize at best on PPTES family -- as depicted in Fig.~\ref{fig:generalization}.

\section*{Quantum entanglement}
Before discussing the introduced methods for the identification of entanglement in a given $n$-qubit quantum system, let us introduce basic concepts and definitions used when discussing this phenomenon. The foundations of quantum mechanics state that in order to describe a system that forms a $n$-qubit register, one must define a $2^n$-dimensional state space (known as the Hilbert space).
The most convenient (and general) way to describe these multi-qubit systems is to construct a density matrix:
\begin{equation}
\rho = \sum_{i} p_{i}\ket{\psi_{i}}\bra{\psi_{i}},
\end{equation}
where $p_{i}$ is a probability that the system is in a given state $\ket{\psi_{i}}$. In this general approach, a quantum system is represented by a mixture of states, or in other words a mixed state. However, if only one state is used to determine the whole system, then it is said, that it is a pure state. In such case, the density matrix is reduced to the outer product of a state vector $\ket{\psi}$  with itself $\rho = \ket{\psi}\bra{\psi}$.

Having a complex system, i.e., a system consisting of at least 2 qubits, one can define a separable state as the one, that can be separated to a tensor product of sub-system states. 
This definition is valid in the case of pure states, however, 
is has to be extended to capture mixed states.
In the latter case, the bi-separable mixed state can be written as:
\begin{equation}
    \rho_{AB} = \sum_i p_i \rho_A^i \otimes \rho_B^i,
\end{equation}
where $\sum_i p_i = 1$, while $\{\rho_A^i\}$ and $\{\rho_B^i\}$ are the density matrices representing states of the corresponding sub-systems $A$ and $B$. If the mixed state cannot be represented as such a convex sum ($p_i \ge 0$) of product states, then the state is entangled.

In general, determining the separability of a given state using the above-mentioned definition is a non-trivial problem.\cite{zyczkowski2017} Therefore, many different criteria have been constructed to decide whether a state is entangled.
One noteworthy example is the negativity\cite{zycz1998} metric, which is defined as follows:
\begin{equation}\label{eq:neg}
    Neg( \rho_{AB}) = \sum_{\lambda_i < 0} |\lambda_i|,
\end{equation}
where $\lambda_i$ are the negative eigenvalues of the partially transposed density matrix $\rho^{T_B}$ associated with the given bipartition $A|B$ of the whole system. The partial transpose operation for its density matrix $\rho_{AB} = \sum_{i,j,r,s}a_{i,r}a^*_{j,s}\ket{i}_{A}\bra{j}_{A}\otimes\ket{r}_{B}\bra{s}_{B}$, with respect to the subsystem $B$ is defined as:
\begin{equation}
\begin{split}
\rho^{T_B} = \sum_{i,j,r,s}a_{i,r}a^*_{j,s}\ket{i}_{A}\bra{j}_{A}\otimes(\ket{r}_{B}\bra{s}_{B})^T = \\
=\sum_{i,j,r,s}a_{i,r}a^*_{j,s}\ket{i}_{A}\bra{j}_{A}\otimes\ket{s}_{B}\bra{r}_{B}.
\end{split}
\end{equation}

The negativity metric $Neg(.)$ is based on the PPT separability criterion.\cite{Peres, Horodeccy} It states, that non-negative eigenvalues of the partially transposed density matrix are the necessary condition of separability. However, it is a sufficient condition only for $2\otimes2$ or $2\otimes3$ systems. In the case of larger systems, one can be certain that if the negativity is larger than zero (NPT states), then the state is entangled. On the other hand, if the negativity is equal to zero (PPT states) it is not always true that the state is separable, i.e. we may have PPTES. 

There exist some other methods of calculating entanglement, for instance entanglement of formation or entanglement witnesses, however they are far more complex (time\--con\-sum\-ing) because they require optimization in high-dimensional spaces \cite{wooters,Carvalho, PhysRevA.62.052310}, and therefore impractical to use in case of large datasets. All in all, one arrives with analytical methods that are either hard to calculate or sufficient only for pure states or low-dimensional mixed states. On the other hand, there exist numerical algorithms, which are able to qualify the entanglement after a finite time with a desired precision~\cite{geomQE, Doherty_2002, Hulpke_2005}, however, our attempts to use them as \textit{numerical} labels were rather unsuccessful due to a problem with convergence of these algorithms. 
That is why artificial NNs, which are also approximate numerical methods, but statistical in nature, are supposed to cope with this task.   

\section*{Neural network architectures}
Detecting quantum entanglement in many-qubit systems is inseparably connected with analyzing the quantum state of the system, which is represented by a density matrix. 
The most natural concept of a NN designed to process such an input (in the form of a finite two-dimensional array) is a convolutional NN (CNN). 
This type of ML model has already proved its efficiency in dealing with quantum entangled systems.~\cite{levine2019,xiao2021,Choo2019,broecker,roth2021group}. 
In this paper, we propose to use CNN as a feature-extracting backbone followed by FC classification layers in three different configurations presented in Fig.~\ref{fig:arch}.

\subsection*{Deep convolutional networks}
\begin{figure*}[tb]
    \centering
    \includegraphics[scale=0.15]{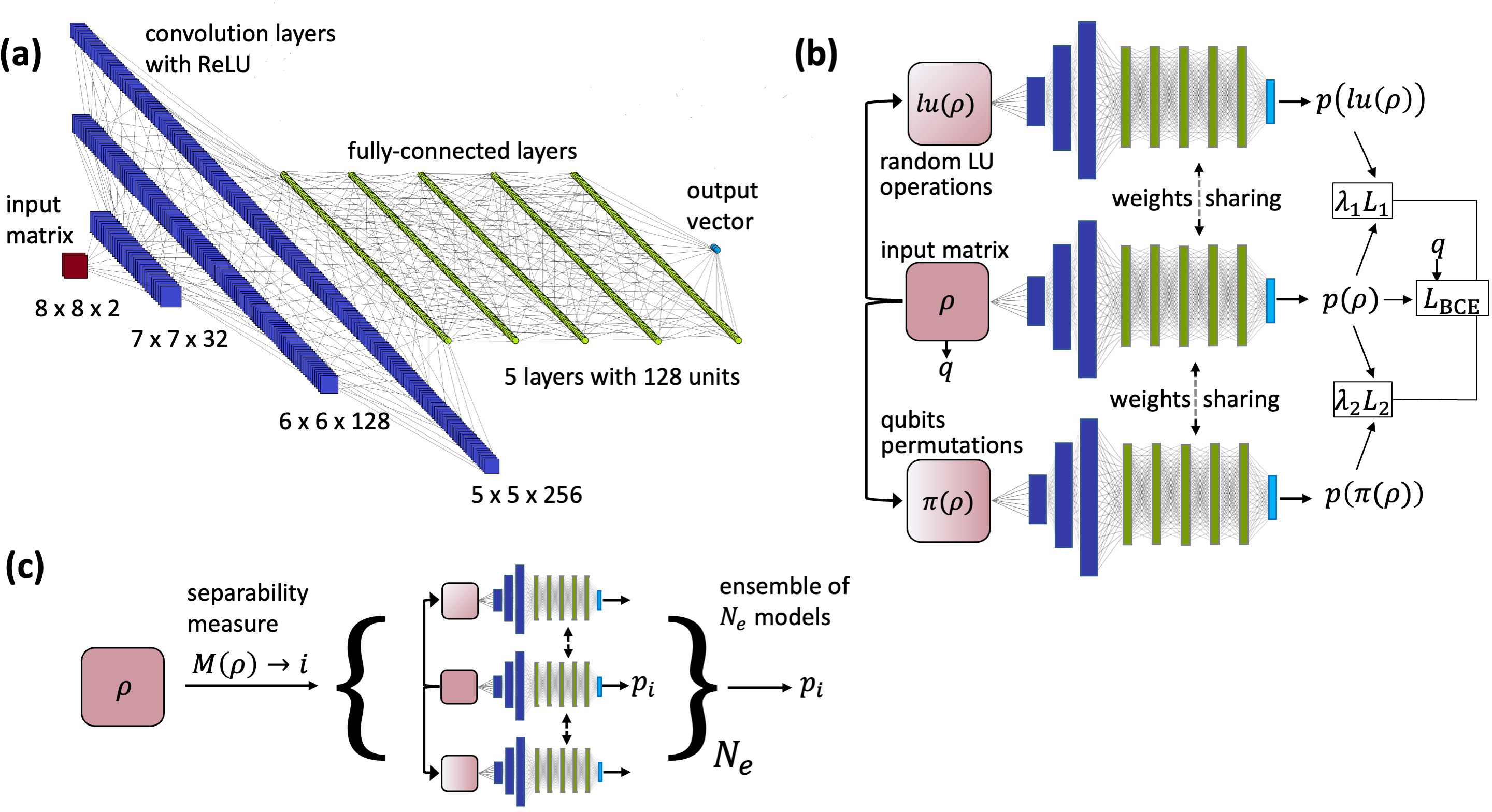}
    \caption{Model architectures: (a) NN for 3-qubit input density matrix (red): CNN backbone with three convolutional layers, each of $2\times2$ kernels (dark blue blocks); five fully-connected layers (green blocks); and finally, 3-element output vector identifying entanglement in three possible bipartitions (light blue).
    (b) Architecture of the triple Siamese network, where each subnetwork, being the CNN from the model (a), is fed by a different input (original or transformed), but shares the same network parameters (weights).
    (c) Ensemble of Siamese networks with separate $N_e$ Siamese models trained on different domains determined using separability measure $M$ (being a pretrained autoencoder model).}
    \label{fig:arch}
\end{figure*}

The first network architecture composed using a single CNN is shown in Fig.~\ref{fig:arch}(a). The input layer has size $K \times K \times 2$, where $K=2^N$ is the Hilbert space dimension and the density matrix is split into the two channels by taking the real and the imaginary part separately. Subsequently, there are three convolutional layers with ReLU activation functions, each with a kernel of size $2 \times 2$, and the number of channels in the $i$-th layer $c_i = \lfloor r_{i}c_{i-1} \rfloor$. Here, the $r_i = \sqrt{r_{i-1}}$ parameter is the layers increasement ratio initialized with $r_1 = 16$, and $\lfloor.\rfloor$ denotes \textit{floor} operation.
Convolutional layers are followed by five fully- connected layers with 128 units each and ReLU activation functions. Finally, $m$-units output vector with sigmoid activation functions is used, where $m$ corresponds to the number of possible bipartitions in the $N$ qubits system.
The architecture of the CNN model for the 3-qubit system is presented in Fig.~\ref{fig:arch}(a), however, this construction can be easily extended to a larger $N$-qubit register.

The network is trained in a supervised manner. To evaluate the loss function we use binary cross entropy averaged over all possible bipartitions:
\begin{equation}
\begin{split}
&\mathcal{L}_\mathrm{CNN}=L_\mathrm{BCE}\left(\boldsymbol{p}(\boldsymbol{\rho}\right),  \boldsymbol{q}) =\\ 
&=\frac{-1}{mn} \sum_j^m \sum_i^n q_{ij} \log(p_j(\rho_{i})) + (1 - q_{ij}) \log(1 - p_j(\rho_{i})),
\end{split}
\end{equation}
where $m$ is the number of bipartitions, $n$ is the batch size, $q_{ij}$ is the label for the $j$-th bipartition of the given state $\rho_i$ (with 0 meaning separable, and 1 -- entangled bipartition), and $p_j(\rho_{i})$ is the predicted probability that $\rho_{i}$ has parts for $j$-th bipartition entangled.

\subsection*{Siamese networks}
Entanglement should be invariant to local operations taken on separate qubits.
 In other words, applying single qubits rotations or phase modification does not influence the entanglement between different qubits. Moreover, permutations of the qubits in the system also should not have any impact on the entanglement identification. Having those two facts in mind, we propose to extend the CNN into a triple Siamese NN, to force the model to be immune to such symmetry operations, and thus improve its robustness. Specifically, in this approach instead of having only one density matrix on the input, we augment it into three matrices, which are processed parallelly. 
 The characteristic is that each subnetwork is an exact copy of the original CNN but with the model weights \textit{shared} between them -- see Fig.~\ref{fig:arch}(b).
 The first input is the original matrix, the second one represents the original state with extra random local unitary operations (LU) applied to each qubit, and the third one corresponds to the density matrix, which is permuted with respect to randomly selected qubits.
 Hence, now the loss for a given batch is defined as follows:
\begin{equation}
\begin{split}
\mathcal{L}_\mathrm{siam} 
&=\lambda_0 L_\mathrm{BCE}\left(\boldsymbol{p}(\boldsymbol{\rho}),\boldsymbol{q}\right) +\\
&+\lambda_1 \frac{1}{mn} \sum_j^m \sum_i^n \left| p_j(\rho_{i}) - p_j(lu(\rho_{i})) \right| +\\  &+\lambda_2 \frac{1}{mn} \sum_{k = \pi(j)}^m \sum_i^n \left| p_j(\rho_{i}) - p_k(\pi(\rho_{i}) \right|,
\end{split}
\end{equation}
where by $lu(\rho_{i})$ we denote some random LU operation on state $\rho_i$, $\pi(\rho_i)$ is the permutation of state $\rho_i$ chosen independently for a given batch, and $\lambda_1$, $\lambda_2$ are hyperparameters which allow to adjust the influence of such extensions of the loss function.
At the beginning $\lambda_0=1$, $\lambda_1=0$ and $\lambda_2=0$,  while during the training after 10 epochs, we increased the $\lambda_2$ and $\lambda_3$ regularizers to $0.5$. 
This shifts the impact from label-based part of the loss (controlled by $\lambda_0$) to symmetrizing (i.e., enforcing symmetries) part ($\lambda_1$ and $\lambda_2$), which is a typical approach for \textit{semi-supervised} training schemes.
One important remark is, that permutation $\pi$ can rearrange the output vector in case of certain bipartitions. 
For example, 3-qubit permutation $\pi = (1,3,2)$ does not change the predicted probability of entanglement for the bipartition (1|23). Contrarily, the same permutation switches the probabilities for the other two bipartitions (2|13 and 3|12).
Therefore, corresponding indexes must be selected carefully, which we denoted by writing $k=\pi(j)$.   

\subsection*{Ensemble of networks}
\begin{figure}[!b]
    \centering
    \includegraphics[scale=.45]{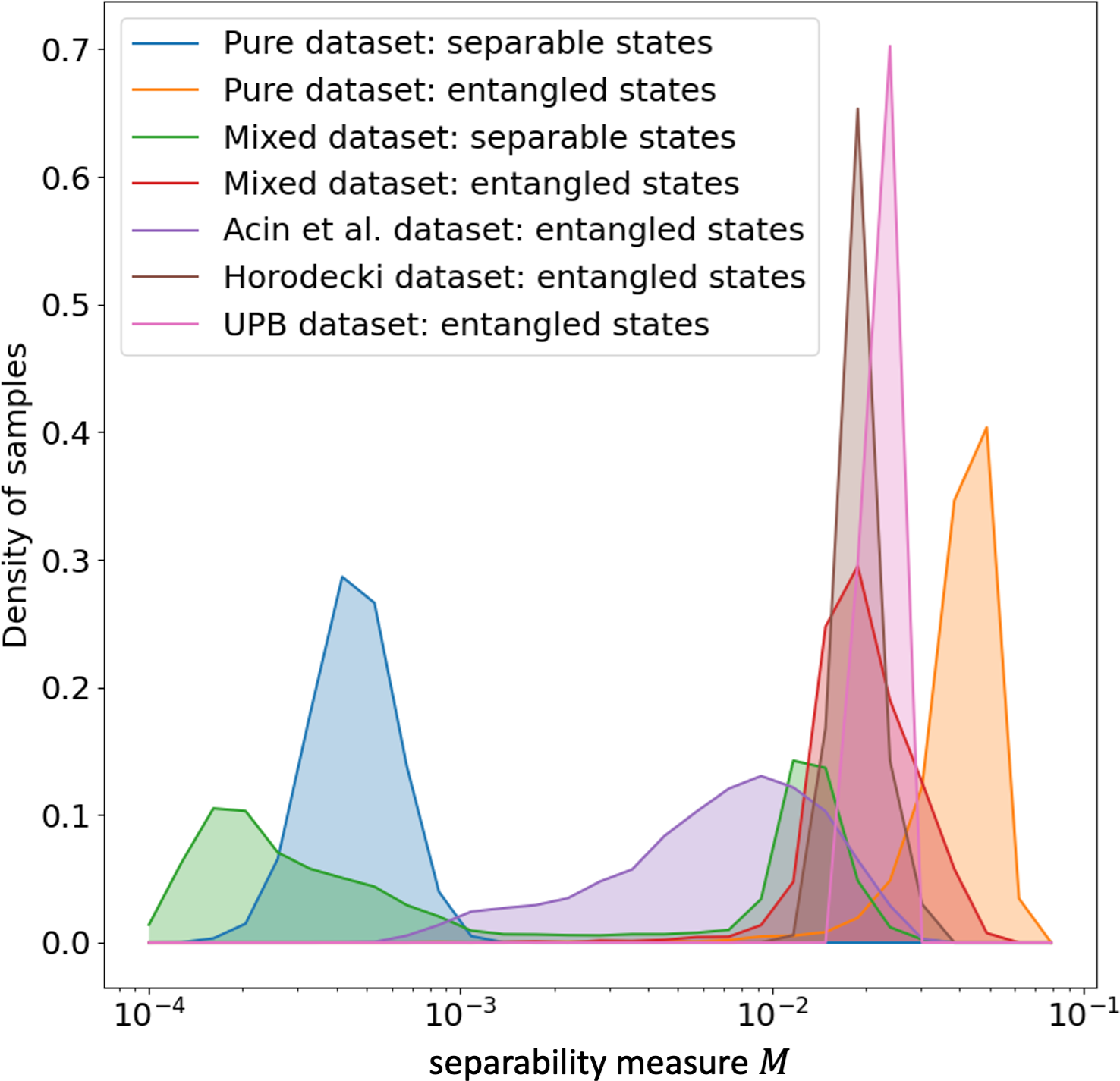}
    \caption{Idea of the ensemble method: various states can be distinguished on the basis of some external measure (here defined by a pretrained autoencoder loss $M$) and then learned separately by \textit{ensemble} of models that work independently in their $M$ domains.}
    \label{fig:ensemble}
\end{figure}
During the experiments, we observed that there is a significant difficulty in forcing NN models to recognize mixed separable and entangled states at the same time. Thus, to cope with this issue we decide to train several Siamese models, forming \textit{ensemble} of size $N_e$, that work separately on their domains -- see scheme in Fig.~\ref{fig:arch}(c).
Different classes of states falling into separate domains can be distinguished on the basis of some external measure $M$ that quantifies their separability. 
To obtain such a measure, we use a pre-trained autoencoder NN model, introduced in our recent work [\onlinecite{discord}]. The autoencoder was trained to reconstruct separable mixed states, hence the reconstruction loss can measure whether a given state is similar to separable states, and consequently serve as machine-learned separability measure $M$. Having values of $M$ for different states, we form $N_e = 10$ domains with logarithmic intervals $[0, 0.0001)$, $[0.0001, 0.0002)$, $[0.0002, 0.0005)$, $[0.0005, 0.001)$, $[0.001, 0.002)$, $[0.002, 0.005)$, $[0.005, 0.01)$, $[0.01, 0.02)$, $[0.02, 0.05)$ and $[0.05, \infty)$.

In figure Fig~\ref{fig:ensemble} there are presented distributions of samples from various subsets, appearing in the training and test sets, according to the value of separability measure. It is seen that the chosen measure separates well the separable and entangled states, apart from some specific group of mixed separable states, which are confused by the network with entangled states. In fact, if this subset had not existed, one could have simply utilized the unsupervised autoencoder as the perfect tool to distinguish between separable and entangled states, as proposed in Chen et al.~\cite{Chen2021} However, with the current approach of training supervised models specifically on the subsets of the data set, we can force the networks to learn the characteristic patterns of the given subgroups, and therefore allow them to detect states omitted by the autoencoder. One can notice, that it is in full agreement with the analytical approach of finding separate types of entanglement witnesses for different classes of states. 

What may also be beneficial, the various PPTES classes are separated somewhat with respect to the $M$ measure (cf.~Fig~\ref{fig:ensemble}). Interestingly, the distribution of the Acin et al. states seems to be the closest to the separable states, hence correct classification of these states may be the most problematic.

Nevertheless, due to its specification, the Ensemble model should be able to recognize entanglement patterns with higher accuracy than standard models and generalize well on PPTES states, which were not present in the training datasets as shown on the scheme in Fig.~\ref{fig:generalization}.

\section*{Training data generation}
The most crucial thing to train a ML model is to have well-balanced and diversified dataset. Thus, in order to achieve this goal, we propose various methods of density matrices generation, that will be used to build synthetic datasets for training our NN models.

\begin{figure*}[bt]
    \centering
    \includegraphics[scale=0.55]{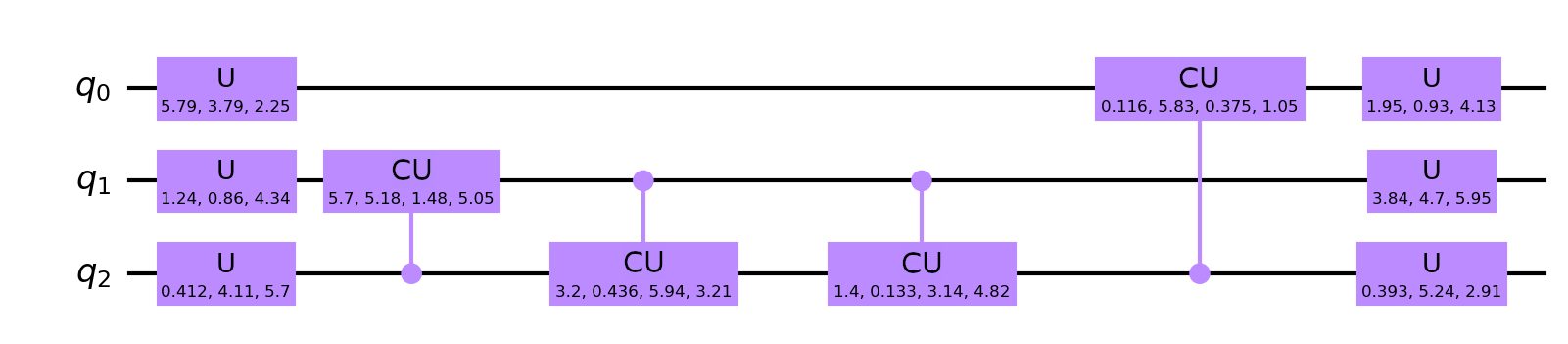}
    \caption{Example of a circuit generating a 3-qubit random state. It consists of $U$ and $CU$ gates (violet blocks) with random parameters written in each block (and defined in the Appendix).}
    \label{fig:rand}
\end{figure*}
Two main techniques were used to generate pure random states. The first idea is to use a quantum circuit model that contains local single-qubit gates $U$ that randomize separate qubits, and the random number of two-qubit controlled gates $CU$, that can entangle qubits. The $CU$ gates are put into random pairs of qubits. 
Combining these two types of quantum gates with their appropriate parameters (angles), one can obtain a suitable generator of pure states, both entangled and separable ones.
The detailed parametrization of the gates and circuit organization can be found in the Appendix.
An example of a 3-qubit circuit generated by this method is presented in Fig.~\ref{fig:rand}.
The second idea of generating random pure states is based on sampling from the uniform Haar measure~\cite{Haar} described in the Appendix.

Having these two methods, one can suspect that using both of them produces a sufficiently diversified dataset of the pure state vectors and consequently density matrices. However, apart from them, we also included three specific types of quantum states, which may hardly occur when using both procedures.
Namely, using the circuit approach we intentionally generate pure separable states (here $CU$ gates are absent in the circuit), the GHZ state, and the W state~\cite{cruz2019}, each of them being locally randomized with the $U$ gates. 
It is obvious, that having only pure states included in the training dataset may result in missing the possibility of detecting mixed states. That is why, we should also think of generating such states straight from the definition of a mixed state: 
\begin{equation}
\label{eq:mixed_def}
\rho = \sum_{i}^{d} p_{i}\ket{\psi_{i}}\bra{\psi_{i}},
\end{equation}
where for training purposes we select $d$ at random over the interval $[2, 2^{N}]$, and $N$ is a number of qubits.
In this na\"ive method one simply mixes many pure states $\ket{\psi_i}$ with random probabilities $p_i$. However, to have as diversified dataset as possible, we also use the second method of generating mixed states:
\begin{equation}
\label{eq:mixed_tr}
\rho_{AB} = \mathrm{tr}_{C}(\ket{\psi_{ABC}}\bra{\psi_{ABC}}),
\end{equation}
with a partial trace operation $\mathrm{tr}_C(.)$.
Namely, we generate a larger pure system $\ket{\psi_{ABC}}$ and trace out some part of it (subspace $C$), until the desired size of the system $AB$ is achieved. This way, if the global space $ABC$ is entangled, one can be sure that the reduced density matrix $\rho_{AB}$ is mixed.

To train a NN in a supervised learning scheme one needs to have labels $q$ -- see Fig.~\ref{fig:arch}(b), which classify
each bipartition of the system in a given state as entangled or separable. Then, each state $\rho_i$ is labeled by a binary vector $q_{ij}$, where subsequent elements $j$ encode the presence of entanglement for the given bipartition.
In the case of pure states, we calculate the negativity measure (Eq.~\ref{eq:neg}) for each bipartition and mark it accordingly to its value: if $Neg=0$, it is marked as separable, while if $Neg>0$ it is marked as entangled. In the case of mixed states, where it is not guaranteed that the negativity measure distinguishes separable and entangled states, we either generate states that are separable (by definition) or entangled (by taking only NPT, i.e. where the negativity is conclusive), or in case of PPT we propose two different strategies for labeling unknown states. This latter group is important because it contains PPTES, and therefore allows us to train a model to capture them.

The most na\"ive strategy is based on negativity metric and simply labels mixed states accordingly to the $Neg$ value -- we will call this strategy as \textit{negativity labeled}.
However, the main problem with this approach is that it may falsely label mixed PPT states, i.e. where the criterion is inconclusive. That is why we introduce a second strategy called \textit{verified}, where the 
PPT states are simply excluded from the data.
The main disadvantage of such a solution is that it automatically excludes all of mixed separable states. To minimize this drawback, we supplement the dataset by generating them straight from the definition.
Further descriptions of the labeling strategies can be found in the Appendix.

\begin{table*}[bt]
\centering
\begin{tabular}{|c|l|c|l|}
\hline
\textbf{state type}                     
&\multicolumn{1}{c|}{\textbf{generating method}}
& \textbf{labeling method}              
& \textbf{amount} \\ 
\hline
pure separable  
& \begin{tabular}[c]{@{}l@{}}circuit with random 1-qubit gates only \\ 
(separable circuit)\end{tabular}
& negativity                            
& 40 000 \\
\hline
pure biseparable  
& \begin{tabular}[c]{@{}l@{}}Kronecker product of 1-qubit state and 2-qubit state, \\  
sampling from Haar measure\end{tabular}
& negativity                            
& 20 000 \\ 
\hline
pure entangled  
& \begin{tabular}[c]{@{}l@{}}random circuits, including W and GHZ, \\ and sampling from the Haar measure\end{tabular}
& negativity                            
& 60 000 \\ 
\hline
mixed separable 
& \begin{tabular}[c]{@{}l@{}l@{}}separable circuit, \\ 
Kronecker product (\ref{eq:mixed_kron}) of random 1 qubit states, \\
and Kronecker product (\ref{eq:mixed_kron}) of traced 1 qubit states\end{tabular}
& by definition 
& \begin{tabular}[c]{@{}l@{}l@{}}20 000 +\\ 
120 000 + \\
20 000\end{tabular} \\
\hline
mixed biseparable 
& Kronecker product (\ref{eq:mixed_kron_bisep}) of 1 qubit state, and 2-qubit state
& \begin{tabular}[c]{@{}c@{}}2 strategies:\\ \textit{negativity}, \textit{verified}\end{tabular} 
& \begin{tabular}[c]{@{}l@{}}120 000\end{tabular} \\
\hline
mixed entangled 
& \begin{tabular}[c]{@{}l@{}}from the definition (\ref{eq:mixed_def}), \\ and using partial trace (\ref{eq:mixed_tr})\end{tabular} 
& \begin{tabular}[c]{@{}c@{}}2 strategies:\\ \textit{negativity}, \textit{verified}\end{tabular} & \begin{tabular}[c]{@{}l@{}}100 000 +\\ 
60 000\end{tabular} \\ 
\hline
\end{tabular}
\caption{Training datasets composition. Both sets have the same structure of pure and mixed separable states but have different strategies for labeling mixed entangled states.}
\label{tab:data}
\end{table*}
Having defined various methods for the generation of pure and mixed states together with two different strategies for labeling mixed states we have generated training datasets. They are summarized in Table~\ref{tab:data}, along with the number of generated states for each group. In total, we have synthesized two training sets, each with the same structure of the pure (120~000 states) and mixed separable (160~000 states), but with different strategies for labeling mixed entangled (160~000 states) and mixed biseparable (120~000 states). 

In order to correctly evaluate the model, apart from the training sets, the validation and test sets were also created.
The first validation set is generated with the same structure as the \textit{verified} training dataset but with 10 times smaller number of examples, i.e. 56~000 states generated. Additionally, the second validation data set consisting of 10~000 $2 \otimes 2^{N-1}$ dimensional PPTES states \cite{tu2012} is generated as well. Next, we defined two test datasets for pure (30~000 states) and mixed states separately. 
The latter contains 20~000 separable states, and 20~000 entangled containing NPT states only (\textit{verified} strategy), generated both from the definition (Eq.~\ref{eq:mixed_def}) and using partial trace (Eq.~\ref{eq:mixed_tr}). Therefore, the test dataset contains only the correctly labeled states. Further details on states generation can be found in the Appendix.

Since the main reason for creating the artificial classifier is to have the ability to detect not only NPT entangled states but also PPTES, we generate 3 separate test datasets, each containing 10 000 well-defined PPTES of different types. Precisely, the first PPTES test set includes $2 \otimes 4$ PPTES introduced by Horodeccy\cite{horodecki1997} and then generalized to $N$-qubits, $2 \otimes 2^{N-1}$ dimensional states\cite{tu2012}. The second test set is generated with the use of unextendible product basis (UPB)\cite{be1999}, with a similar approach being described by Ma et al.~\cite{ma2018} The third one is composed of fully PPT entangled states proposed by Acin et al.\cite{acin2001} 
All of these states are further randomized by applying local operations to the separate qubits.

\section*{Results}
After we have defined the deep NN models and generated the labeled training and test sets, we can move on to training and then testing our models.
To verify the models' fidelity we used the standard accuracy metric, which is defined as follows:
\begin{equation}\label{eq:acc}
    Acc = \frac{1}{mn}\sum_{i=1}^m TP_i + TN_i,
\end{equation}
where $m$ is the number of possible bipartitions, $TP_i$ and $TN_i$ represent the number of true positive (entangled classified as entangled) and true negative (separable classified as separable) examples for a given bipartition $i$, and $n$ is the total number of samples in the test datasets. 

\begin{table*}[bt]
\vspace{5mm}
  \centering
    \begin{tabular}{c|c|c|c|c|c}
    \rowcolor[rgb]{ .859,  .859,  .859} \textbf{Models} & \cellcolor[rgb]{ 1,  .902,  .6}\textbf{Pure states} & \cellcolor[rgb]{ .776,  .878,  .706}\textbf{Mixed states} & \cellcolor[rgb]{ .8,  .6,  1}\textbf{Horodecki PPTES} & \cellcolor[rgb]{ .741,  .843,  .933}\textbf{UPB PPTES} & \cellcolor[rgb]{ 0,  1,  1}\textbf{Acin PPTES} \\
    \rowcolor[rgb]{ .929,  .929,  .929} CNN   & \cellcolor[rgb]{ 1,  .949,  .8}97.40 & \cellcolor[rgb]{ .886,  .937,  .855}86.39 & \cellcolor[rgb]{ .8,  .8,  1}83.01 & \cellcolor[rgb]{ .867,  .922,  .969}88.85 & \cellcolor[rgb]{ .8,  1,  1}57.45 \\
    \rowcolor[rgb]{ .929,  .929,  .929} Siamese CNN & \cellcolor[rgb]{ 1,  .949,  .8}98.31 & \cellcolor[rgb]{ .886,  .937,  .855}90.79 & \cellcolor[rgb]{ .8,  .8,  1}86.99 & \cellcolor[rgb]{ .867,  .922,  .969}94.89 & \cellcolor[rgb]{ .8,  1,  1}50.81 \\
    \rowcolor[rgb]{ .929,  .929,  .929} Ensemble & \cellcolor[rgb]{ 1,  .949,  .8}98.18 & \cellcolor[rgb]{ .886,  .937,  .855}87.88 & \cellcolor[rgb]{ .8,  .8,  1}88.63 & \cellcolor[rgb]{ .867,  .922,  .969}89.58 & \cellcolor[rgb]{ .8,  1,  1}87.36 \\
    \rowcolor[rgb]{ .929,  .929,  .929} MLP based on measurements  & \cellcolor[rgb]{ 1,  .949,  .8}79,31 & \cellcolor[rgb]{ .886,  .937,  .855}79,31 & \cellcolor[rgb]{ .8,  .8,  1}89,67 & \cellcolor[rgb]{ .867,  .922,  .969}99,81 & \cellcolor[rgb]{ .8,  1,  1}36,30 \\
    \end{tabular}%
  \caption{Averaged bipartite entanglement detection accuracy for the 3-qubit system using different models trained on \textit{verified} dataset.
  The results are presented for test sets composed of pure and mixed states, together with the various classes of PPTES. Last row shows results for MLP model that operates directly on measurements set while not on the (reconstructed) density matrix.}
  \label{tab:nopptes_ext_results}%
\end{table*}%

\begin{table*}[tb]
  \centering
    \begin{tabular}{c|c|c|c|c|c}
    \rowcolor[rgb]{ .859,  .859,  .859} \textbf{Models} & \cellcolor[rgb]{ 1,  .902,  .6}\textbf{Pure states} & \cellcolor[rgb]{ .776,  .878,  .706}\textbf{Mixed states} & \cellcolor[rgb]{ .8,  .6,  1}\textbf{Horodecki PPTES} & \cellcolor[rgb]{ .741,  .843,  .933}\textbf{UPB PPTES} & \cellcolor[rgb]{ 0,  1,  1}\textbf{Acin PPTES} \\
    \rowcolor[rgb]{ .929,  .929,  .929} CNN   & \cellcolor[rgb]{ 1,  .949,  .8}97.37 & \cellcolor[rgb]{ .886,  .937,  .855}88.89 & \cellcolor[rgb]{ .8,  .8,  1}82.80 & \cellcolor[rgb]{ .867,  .922,  .969}95.78 & \cellcolor[rgb]{ .8,  1,  1}5.71 \\
    \rowcolor[rgb]{ .929,  .929,  .929} Siamese CNN & \cellcolor[rgb]{ 1,  .949,  .8}98.43 & \cellcolor[rgb]{ .886,  .937,  .855}91.49 & \cellcolor[rgb]{ .8,  .8,  1}83.04 & \cellcolor[rgb]{ .867,  .922,  .969}87.39 & \cellcolor[rgb]{ .8,  1,  1}1.04 \\
    \rowcolor[rgb]{ .929,  .929,  .929} Ensemble & \cellcolor[rgb]{ 1,  .949,  .8}98.13 & \cellcolor[rgb]{ .886,  .937,  .855}89.71 & \cellcolor[rgb]{ .8,  .8,  1}84.51 & \cellcolor[rgb]{ .867,  .922,  .969}90.61 & \cellcolor[rgb]{ .8,  1,  1}4.01 \\
    \end{tabular}%
  \caption{Test results as in Table~\ref{tab:nopptes_ext_results}, but now the models trained on the \textit{negativity labeled} dataset.}
  \label{tab:negativity_results}%
\end{table*}%

We also optimized the architecture hyperparameters by training the models on the \textit{verified} set and evaluating on the validation sets -- discussion can be found in the Appendix. The results show that the most efficient architecture is the network (CNN or Siamese CNN) with three convolutional layers and a kernel of shape $2 \times 2$.

Let us analyze the results for the smallest configuration where nontrivial entanglement relations may occur, i.e. 3-qubit system.
In Tables~\ref{tab:nopptes_ext_results} and \ref{tab:negativity_results} there are presented results for CNN, Siamese, and Ensemble models, trained on both training sets: 
\textit{verified} and \textit{negativity labeled}. The accuracy was calculated for all generated test sets: with pure states, mixed states, and with various PPTES classes.

When analyzing Table~\ref{tab:nopptes_ext_results}, the first thing, that one can notice, is that all models work pretty well in the case of pure states, reaching accuracies over 97\%. However, as one might expect, the detection of mixed states is slightly weaker.
When it comes to comparing CNN, Siamese CNN, and Ensemble networks one can see the advantage of the Siamese CNN model in better recognizing mixed states than CNN: 91\% vs. 86\%. In the case of PPTES, we also observe 
the advantage of using the Siamese model which gives better for Horodecki (87\% vs. 83\%) and UPB states (95\% vs 89\%). However, in the case of Acin states the fidelity for the Siamese network is slightly lower (51\% vs. 57\%) -- we suspect that this is due to the fact that Acin states are `closer' to separable states than two other PPTES (cf. Fig.~\ref{fig:ensemble}), causing the Siamese detector (which is more sensitive to PPTES) to generate more false-positive predictions.
It is worth to remind here, that all models were trained on the datasets, which do not include PPTES. This means, that the models can generalize well to Horodecki and UPB states, and slightly worse to Acin states.

\begin{figure}[b]
\centering
\includegraphics[width=0.99\linewidth]{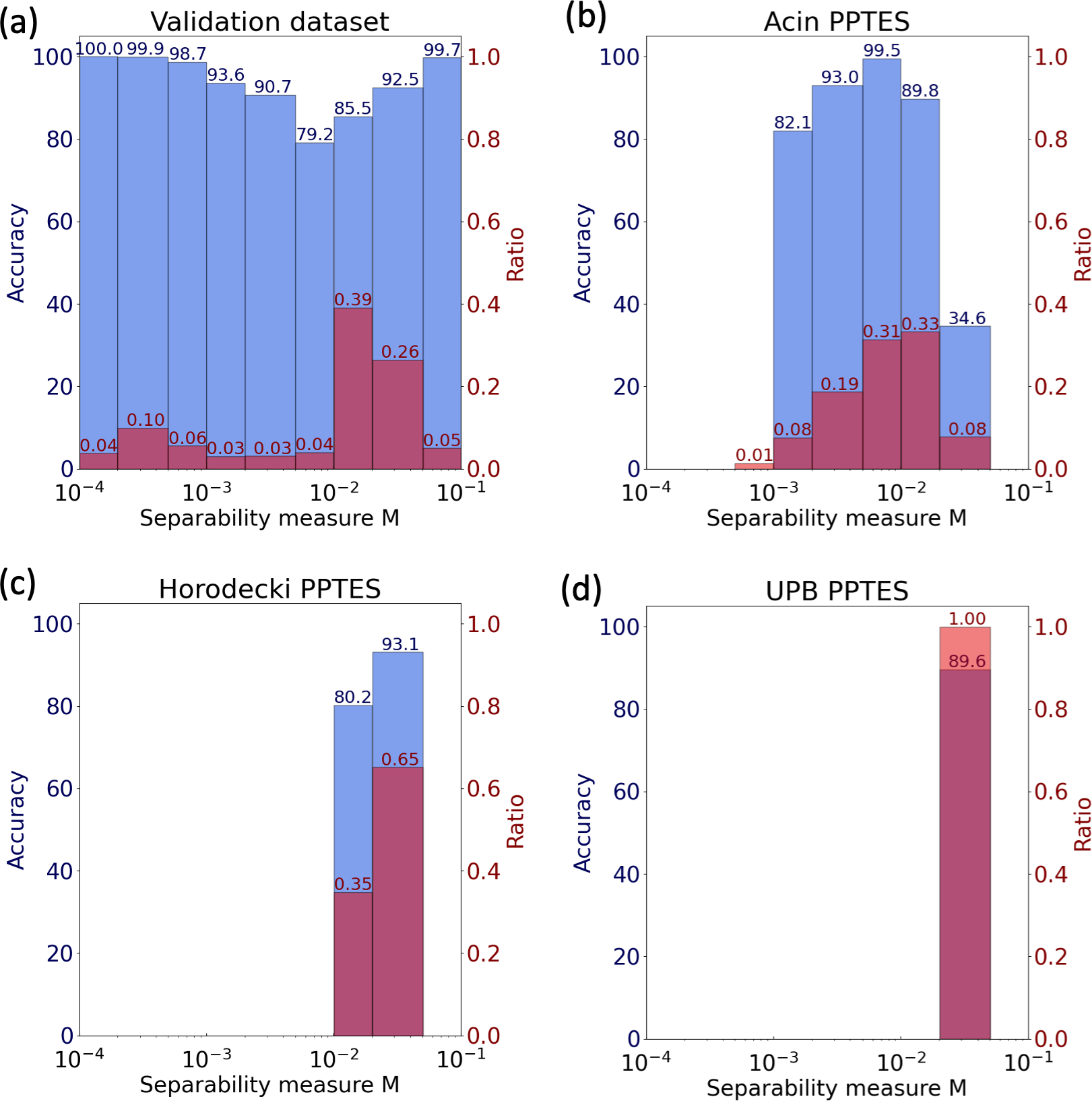}
  \caption{Accuracy (blue bars) of the Ensemble models plotted separately for the subsequent models' domains, determined using the separability measure $M$. Ratio (red bars) characterizes the population of the subsequent domains by the states included in various datasets: (a)~validataion, (b)~Acin PPTES, (c)~Horodecki PPTES, (d)~UPB PPTES. Purple color means that blue and orange bars overlap.}
  \label{fig:domains_accuracy}
\end{figure}

To handle unsatisfactory results for the Acin states, we introduced the ensemble of networks and thus were able to train several models to capture different PPTES classes separately. This approach improved accuracy for Acin states up to 87\% with slightly lower (than for Siamese) accuracy for UPB: 90\%, and slightly better for Horodecki: 89\%. 
In Fig.~\ref{fig:domains_accuracy} we plot accuracies for the subsequent models' domains in the ensemble showing 
that Acin PPTES are quite well detected over almost all covered domains. This proves the usefulness of the ensemble approach where each model specializes in detecting entangled states included only within its own domain.

As the standard representation for the quantum system is a density matrix, hence naturally we proposed to use it as the model input. However, in principle, one could also train the models to operate directly on measurements set. Such an approach may seem beneficial as it reduces the necessity to tomographically reconstruct the density matrix from measurements~\cite{James2001}. On the other hand, it simultaneously limits the models' ability to capture the structural representation of the data, which is contained only in the arrangement of matrix coefficients.
To compare those two approaches we trained a simple MLP, with measurements on the input, getting slightly worse results, as presented in Table~\ref{tab:nopptes_ext_results}(last row), besides UPB states, for which the model works quite well. Therefore, we conclude that using a density matrix is more efficient strategy, especially that one can always either perform the standard tomography~\cite{James2001} before proceeding with NNs, or embed equivalent linear transformation into the NN model, or even train a separate NN to reconstruct $\rho$ from the complete measurements set.

Contrasting the above results with those for \textit{negativity labeled} training set, presented in Table~\ref{tab:negativity_results}, we notice slightly higher accuracy for the general set of mixed states, but worse for the PPTES classes, especially in the Acin states.
However, we still observe some ability to detect Horodecki and UPB states, suggesting that NNs learned to recognize some features of entanglement more general than the negativity criterion itself.

\section*{Discussion}

The main problem with building a general classifier of entanglement in quantum states is a lack of reliable methods for labeling mixed states. If we label some states with mistakes, the ML methods may learn from the mistakes as well. Thus, while it is good to see the high accuracy of the models for pure or mixed NPT states, PPTES are our main concern. 

However, we were able to show that if we use a properly diversified training dataset, we can force the model to generalize correctly to capture PPTES even if they were not present in the training set. This effect can even be strengthened if we use the architecture of Siamese CNN trained in a semi-supervised scheme with regularization terms that force the model to respect the system symmetries, that do not change the entanglement configuration (LU or respective permutations).

Improved generalization efficiency through the use of Siamese networks shows the possible direction for the development of ML models for describing physical systems with their inherent symmetries. 

Unfortunately, Siamese CNN performs worse with a certain group of states that are inherently closer to separable states, i.e., the Acin states. To overcome this, we introduced the Ensemble model, which is combined with a separability measure (unsupervised autoencoder) that enables the training of multiple models tuned to different groups of PPTES, and is capable of learning the characteristic patterns of the given subgroups. The autoencoder itself, which is trained in an unsupervised way (on separable states), does not fully distinguish between mixed separable and entangled states, but extending it with an ensemble of supervised models leads to synergistic improvement.
There are no PPTES in the training set (training is PPTES-agnostic), thus we can suspect that the Ensemble model should give equally good results for any other type of PPTES. 

During the experiments, we also tested other NN architectures. However,  both generic multi-layer FC networks, as well as more advanced tensor networks~\cite{tensornetwork}, used as a backbone model instead of CNN, gave a worse performance. This shows that convolutional networks can be well suited to efficiently represent entanglement in multi-qubit systems. 

To verify whether the proposed approach can be scaled to larger multi-qubit systems we calculate results presented in Table~\ref{tab:scaling}, showing that the Ensemble model works correctly also for detecting bipartite entanglement in a 4-qubit system. Accuracy for the detection of pure states is only slightly worse, while for mixed states is better, which should be taken with caution. Particularly, if we look at the accuracy for the training set we notice that it is also higher in the 4-qubit case, which misleadingly implies that identifying the entanglement is easier in a larger space. In fact, it simply means that the generated dataset does not cover the complexity of this space (especially in the case of mixed states), and future works should focus on improving the generation methods of mixed states in higher dimensional space. Nevertheless, the model architecture seems to scale correctly.
\begin{table}[b]
  \centering
    \begin{tabular}{c|c|c|c}
    \rowcolor[rgb]{ .859,  .859,  .859} \textbf{System size} & \multicolumn{1}{l|}{\cellcolor[rgb]{ .6,  .6,  1}\textbf{Train dataset}} & \multicolumn{1}{c|}{\cellcolor[rgb]{ 1,  .902,  .6}\textbf{Pure states}} & \multicolumn{1}{c}{\cellcolor[rgb]{ .776,  .878,  .706}\textbf{Mixed states}} \\
    \rowcolor[rgb]{ .929,  .929,  .929} 3 qubits & \cellcolor[rgb]{ .8,  .8,  1}92,29 & \cellcolor[rgb]{ 1,  .949,  .8}98,18 & \cellcolor[rgb]{ .886,  .937,  .855}87,88 \\
    \rowcolor[rgb]{ .929,  .929,  .929} 4 qubits & \cellcolor[rgb]{ .8,  .8,  1}96,28 & \cellcolor[rgb]{ 1,  .949,  .8}97,65 & \cellcolor[rgb]{ .886,  .937,  .855}91,58 \\
    \end{tabular}%
  \caption{Scaling properties of the Ensemble model: comparison of detection accuracies between 3- and 4-qubit systems.}
  \label{tab:scaling}%
\end{table}%


Entanglement is an important feature of quantum systems with huge application potential in quantum computation. Controllable entangled states are resources for quantum communication, quantum teleportation, or quantum key distribution protocols (quantum cryptography). Not all quantum states are equally valuable as a resource~\cite{Chitambar2019}. To quantify this value, different entanglement measures can be used. In real situations, entanglement measures are difficult to compute for an arbitrary mixed state (in a situation when a state is a part of a larger system) as the dimension of the entangled system grows. From this perspective, searching for more reliable and scalable methods of entanglement identification in quantum resources is desirable, in particular usage of neural network-based representations seems to be an interesting method for developing new entanglement detectors.

\section*{Methods and code availability}
To implement the quantum circuit model,
during the training dataset generation, the python Qiskit \cite{Qiskit} library is used.
Also, when sampling from the uniform Haar measure during generation we use the built-in Qiskit method.
All the NN models were implemented, trained, and evaluated using the PyTorch\cite{pytorch} library that supports autograd mechanism.
The typical training time depends on the model size and ranges from 1~hours for the pure CNN model to 3-4~hours for the Siamese CNN and 20 hours for the Ensemble model.
To train the NN models we used Nvidia RTX~4090 GPU.
The whole dataset generation lasts about 1~day for 3-qubit density matrices, 
using a typical few-core CPU.
The code with the neural models' definition and detailed training schemes, as well as the data generation procedures, are available at \href{https://github.com/Maticraft/quantum_correlations}{github.com/Maticraft/quantum\_correlations}.

\section*{Acknowledgements}
Authors acknowledge support from National Science Centre, Poland, under grant no. 2021/43/D/ST3/01989.

\bibliography{main}

\setcounter{figure}{0}
\renewcommand{\thefigure}{A\arabic{figure}}
\appendix*
\section*{Appendix}

\subsection{Circuit for random states generation}

Here we present two methods for generating random pure states.

The first idea is to use a quantum circuit model. It is a common technique in quantum computing when one needs to process states of many qubits. In this approach, firstly we initialize the set (register) of $N$ qubits in a given state (usually $\ket{0}^{\otimes N}$), and then apply so-called quantum gates. One can distinguish two main types of such operations: the local -- single-qubit gates, and the two-qubit controlled gates. Although both of them can be understood as rotations on the qubit sphere by the given angles, in the latter such operation is performed only on the single qubit while the rest of the qubits can be used to determine (control) whether this operation should or should not be performed. 
A controlled gate can be used to entangle two qubits.
In contrast, single qubit gates can only modify the local states of qubits and not influence the global entanglement properties. Therefore, by combining these two types of quantum gates with appropriate parameters (angles), one can obtain a suitable generator of pure states, both entangled and separable ones.
The universal single qubit gate is defined as follows:
\begin{equation}\tag{A1}
\label{eq:u}
    U(\theta, \phi, \lambda) = \left( \begin{array}{cc} cos\frac{\theta}{2} & -e^{i\lambda}sin\frac{\theta}{2}  \\ e^{i\phi}sin\frac{\theta}{2} & e^{i(\phi + \lambda)}cos\frac{\theta}{2}  \end{array} \right),
\end{equation}
where $\theta$, $\phi$ and $\lambda$ are three Euler angles. Subsequently, the controlled universal gate is described as:
\begin{equation}\tag{A2}
\label{eq:cu}
    CU(\theta, \phi, \lambda, \gamma) = \left( \begin{array}{cccc} 1 & 0 & 0 & 0 \\ 0 & 1 & 0 & 0 \\ 0 & 0 & e^{i\gamma}cos\frac{\theta}{2} & -e^{i(\gamma + \lambda)}sin\frac{\theta}{2} \\ 0 & 0 & e^{i(\gamma + \phi)}sin\frac{\theta}{2} & e^{i(\gamma + \phi + \lambda)}cos\frac{\theta}{2} \end{array} \right),
\end{equation}
with additional global phase $e^{i\gamma}$. To obtain the most possible diversified dataset, the parameters of these two gates are chosen at random, where the range for each parameter $\theta$, $\phi$, $\lambda$, and $\gamma$ is equal to $[0, 2\pi)$. 

Let us now describe the circuit organization.
At first, the system is initialized in the random separable state by applying the $U$ gates to each of the qubits in the register.
Next, the random number of $CU$ gates is picked from the range $[1, 2\binom{N}{2})$, where $N$ is the number of qubits.
The $CU$ gates are put to random pairs of qubits.
And finally, the state is locally randomized, again, by applying the $U$ gates.

The second idea of generating random pure states is based on sampling from the uniform Haar measure.
At first, we generate a vector $\textbf{x}$ of size $K = 2^N$ ($N$ is the number of qubits) with its elements being uniformly sampled over an interval $(0, 1]$.
Secondly, we replace each coordinate $x_i$ with $y_i = -\log(x_i)$, therefore obtaining vector $\textbf{y}$ with logarithmic distribution over an interval $[0, \infty)$.
Next, we generate a $K$-dimensional phase vector $\boldsymbol{\gamma}$, where each coordinate is chosen at random from the range $[0, 2\pi)$.
Finally, we construct a state vector as:
\begin{equation}\tag{A3}
\label{eq:haar}
\ket{\psi_\mathrm{Haar}} = \sqrt{\frac{y_1}{\sum_i y_i}} e^{i\gamma_1} \ket{00...0} + ... + \sqrt{\frac{y_K}{\sum_i y_i}} e^{i\gamma_K} \ket{11...1}.
\end{equation}

\subsection{Strategies for states labeling }
We propose two different strategies for labeling PPT mixed states, i.e. in situations where the negativity criterion is inconclusive.

The most simple solution is to trust the negativity metric and always label mixed state accordingly to the $Neg$ value -- we will call this strategy as \textit{negativity labeled}.
However, such an approach may result in a network trained to reflect just the PPT criterion instead of actual entanglement. That is why we can think of the second method, which simply excludes mixed PPT states (with $Neg=0$).
We will call this strategy as \textit{verified}, because all the labels here are again correct.
The main disadvantage of such a solution is that it automatically excludes all of mixed separable states from the training set. To minimize this drawback, we generate some extra mixed separable states straight from the definition: 
\begin{equation}\tag{A4}
\label{eq:mixed_kron}
    \rho_{abc} = \sum_i p_i \rho_a^i \otimes \rho_b^i \otimes \rho_c^i,
\end{equation}
where one can generate single qubit states $\rho_a^i$, $\rho_b^i$, $\rho_c^i$, and combine them with Kronecker product into 3-qubits mixed separable state $\rho_{abc}$. This approach can be easily generated for $N$-qubit states just by extending the number of single-qubit states. Similarly, we can generate biseparable states as:
\begin{equation}\tag{A5}
\label{eq:mixed_kron_bisep}
    \rho_{abc} = \sum_i p_i \rho_a^i \otimes \rho_{bc}^i,
\end{equation}
where single qubit state $\rho_a^i$ is combined with 2-qubit state $\rho_{bc}^i$. Subsequently, in order to obtain states biseparable in different bipartition qubits are randomly permuted.

\subsection{Generated datasets structure}
Having two strategies for labeling mixed states (labeling pure states is obvious), we consider generating separate training datasets, each corresponding to the different strategy: \textit{negativity labeled} or \textit{verified}. 
All of them retain the same structure of pure states: 40~000 pure separable states, and 60~000 pure entangled states -- generated using random circuits (using $U$ (Eq.~\ref{eq:u}) and $CU$ (Eq.~\ref{eq:cu}) gates) including W and GHZ, and sampling from Haar measure method (Eq.~\ref{eq:haar}). Additionally, 20~000 pure biseparable states are generated by sampling 1 and 2-qubit states from Haar measure and then combining them with the Kronecker product.

In the case of mixed states, we generate 20 000 mixed separable states using separable circuits and formula (Eq.~\ref{eq:mixed_kron}). Further 60~000 separable states are generated as a Kronecker product of mixed single-qubit states obtained with the usage of quantum circuits. A similar approach is used in order to generate another 60~000 states, with the remark that single-qubit states are sampled from Haar measure, and 20~000 states, where single-qubit states are obtained by tracing from the larger system.
Next, we generate 120~000 biseparable mixed states, by creating single and 2 qubit states and then, combining them into 3 qubit states by applying the Kronecker product. The first half of these states is obtained by the usage of quantum circuits and another half is sampled from Haar measure.
Then, we generate 100~000 mixed entangled states by mixing pure entangled states explicitly from the definition (Eq.~\ref{eq:mixed_def}), and keeping only those with $Neg>0$ (NPT).
Finally, we add 60~000 mixed states generated with the use of the partial trace from a global pure entangled state (Eq.~\ref{eq:mixed_tr}).
For the last group in which we have no guarantee that states are either separable or entangled, we apply the two labeling strategies. Additionally, the first training dataset, with the \textit{verified} labeling strategy, is extended with 20~000 of mixed states with maximal ranks calculated as a partial trace of the larger fully entangled global system generated with the quantum circuit. Those extra states are handily labeled as entangled.
 

\begin{figure}[bt]
\centering
\includegraphics[width=0.99\linewidth]{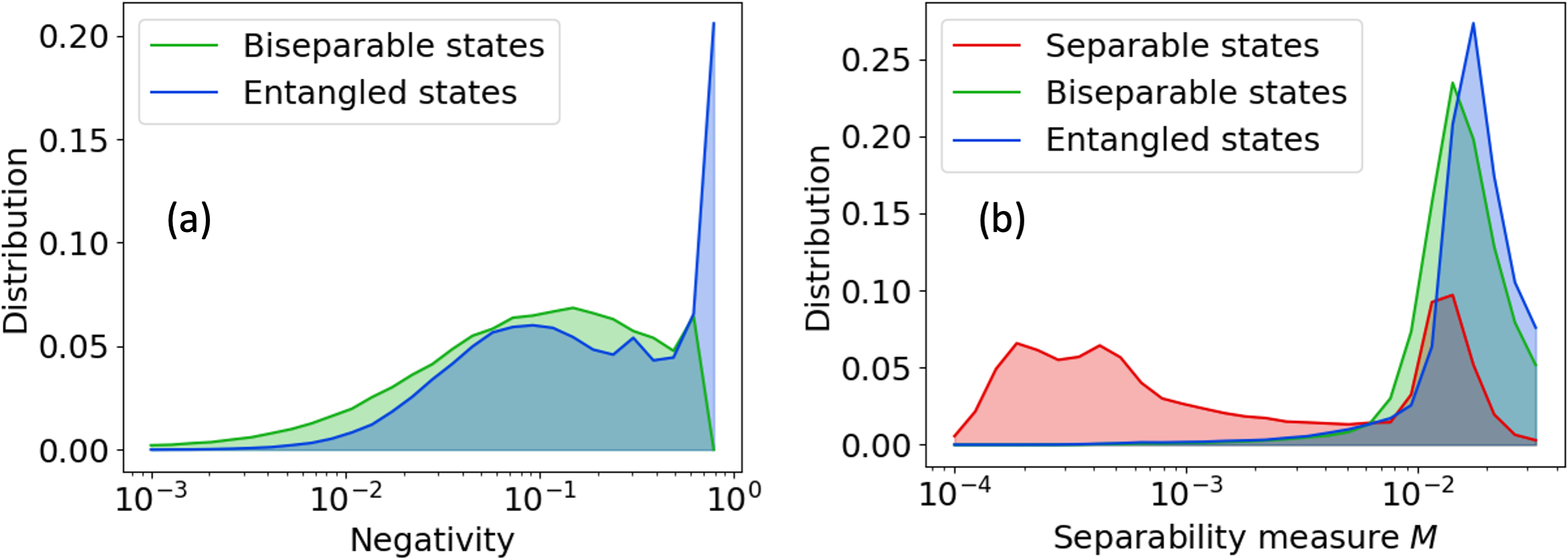}
\caption{Distribution of the training data set plotted against (a) negativity metric and (b) separability measure $M$. Note that for separable states $Neg=0$ by definition, thus they are not visible on the (a) plot.}
\label{fig:train_distribution}
\end{figure}
To verify that generated sets homogeneously cover the areas of 3-qubit space for separable and entangled states, we present Fig.~\ref{fig:train_distribution} showing distributions of above-mentioned groups of states as a function of two measures: negativity and separability $M$. It can be observed that all these three sets fill the corresponding areas fairly homogeneously.

In order to correctly evaluate the model, apart from the training sets, the validation and test sets were also created. The first validation set is generated with the same distribution as the verified training set but with a reduced number of examples, i.e. 10\% meaning 56~000 states generated. The second one consists of 10~000 $2 \otimes 2^{N-1}$ dimensional PPTES states \cite{tu2012}. Next, we define two test datasets for pure and mixed states separately. The former consists of 15~000 separable states, (separable circuits and Kronecker product of single-qubit states),
and 15~000 entangled states (random circuits and sampling from Haar measure). 
The latter contains 20~000 separable states
(mixture of separable circuits and Kronecker product), 
and 20~000 entangled containing NPT states only, generated both from the definition (Eq.~\ref{eq:mixed_def}) and using partial trace (Eq.~\ref{eq:mixed_tr}). 

\subsection{Networks optimization}

Having defined the CNN and the extended Siamese network model, we optimized their architecture hyperparameters (e.g. number of layers or kernel size) by training them on the weakly labeled training set and evaluating on the validation sets. In particular, each model was trained for 20 epochs. Then, we measured the averaged bipartition accuracy on both validation sets and chose the most accurate model.
\begin{figure}[h]
\vspace{5mm}
\centering
\includegraphics[width=0.9\linewidth]{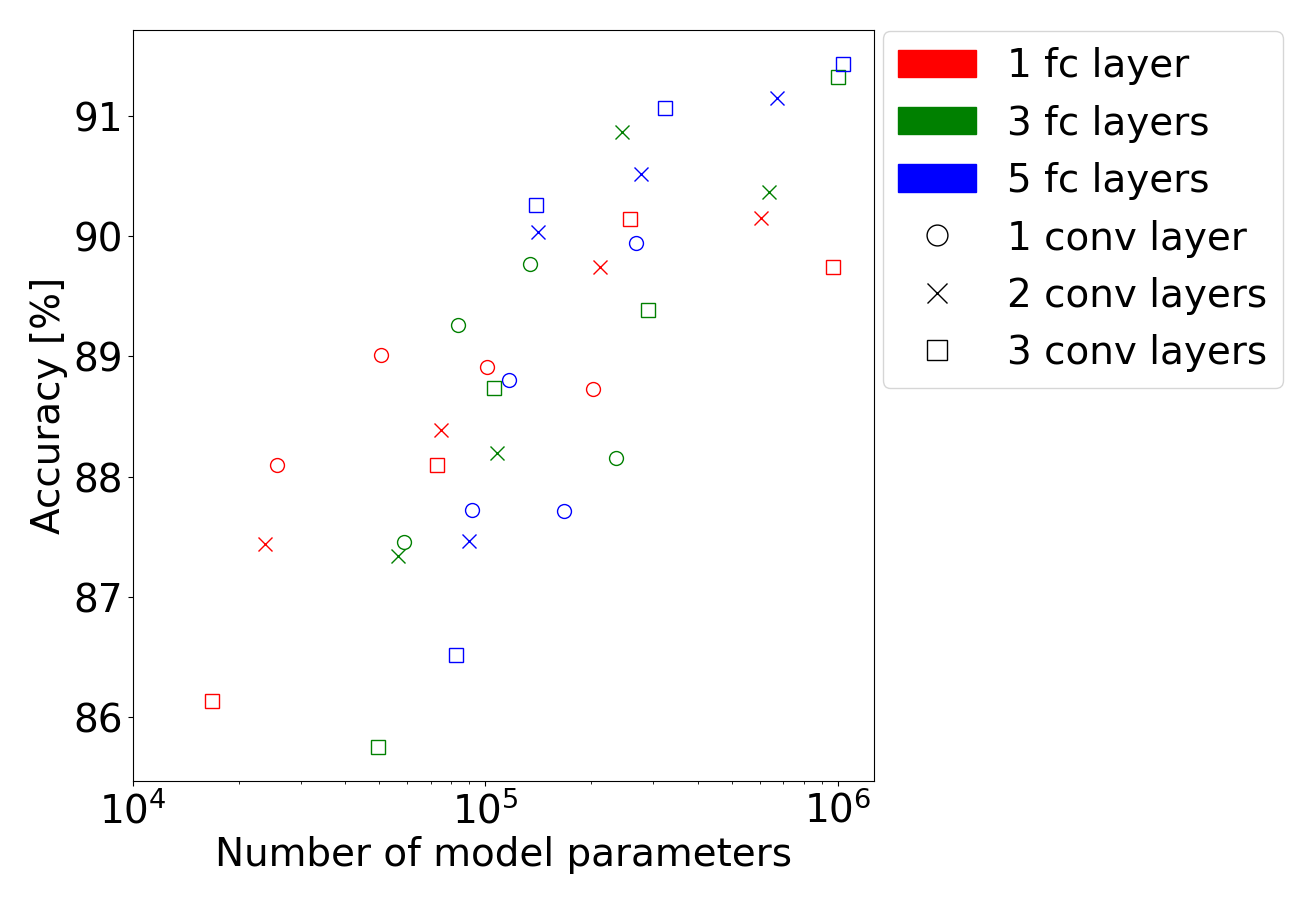}
\caption{CNN model architecture scaling. Number of parameters is increased either by changing the number of fully-connected layers (denoted by different colors), convolutional layers (different marker types) or directly by increasing the number of convolutional channels. The accuracy is measured on the first validation dataset.}
\label{fig:opt}
\end{figure}

Having performed the above procedure for different models, we can compare their architectures and consequently find the most optimal one. This is demonstrated in Fig.~\ref{fig:opt}, where the accuracy for the CNN network models is plotted against the number of parameters. One can notice that using deeper (consisting of more convolutional and FC layers) models tend to have better accuracy. However, as the complexity of the networks increases, by adding more convolutional and FC layers, we observe a saturation in prediction accuracy for the models containing $\sim10^6$ parameters. This is why we decided to limit our architectures, presented in Fig.~\ref{fig:arch}, to 3 convolutional and 5 FC layers only.

It is worth mentioning that the parameter count significantly exceeds the number of linearly independent real parameters needed to characterize a 3-qubit system, i.e., $8^2-1=63$. This is due to a fact that modern NNs are usually \textit{overparameterized}, that is, unlike traditional statistical/ML models where taking too complex models results in overfitting (the so-called bias-variance tradeoff), in DL models the usage of more parameters even improves their properties~\cite{Belkin_2019}. It is then observed as a ``double descent'' curve that generalizes the standard ``U-shaped'' bias-variance curve by showing that increasing model capacity beyond some critical point results in improved performance. Increasing the size of the model can act as a kind of regularization. This suggests that parameter counting cannot indicate the true complexity of deep NNs.

\end{document}